\documentclass[a4paper,11pt]{article}

\newif\ifpdf

\ifx\pdfoutput\undefined
 
\pdffalse

\else
 
\pdfoutput=1
 
\pdftrue

\fi

\ifpdf
 
\usepackage[pdftex]{graphicx}
 
\else
 
\usepackage{graphicx}

\fi

\title{\textbf{Status of the NA61 (SHINE)  experiment at CERN}}

\author{Magdalena Posiadala\footnote{email adress: Magdalena.Posiadala@fuw.edu.pl} \\  for NA61 (SHINE) Collaboration \\ \textit{University of Warsaw, Warsaw, Poland } }

\begin{document}
\maketitle
\begin{center}
\textbf{Abstract}
\end{center}
A precise measurement of the hadron production from interactions of \mbox{30 GeV}
protons on carbon target is performed by NA61 (SHINE)
experiment at the CERN SPS.
The inclusive spectra of pions and kaons on the carbon target obtained from NA61 measurements constrain the neutrino flux in the T2K long baseline
neutrino oscillation experiment at \mbox{J-PARC}, Japan.
The article presents description of the NA61 apparatus together with the preliminary results from the pilot 2007 run.

\newpage

\begin{center}
The NA61 (SHINE) Collaboration 
\end{center}

\noindent N.~Abgrall${}^{22}$,
A.~Aduszkiewicz${}^{23}$,
B.~Andrieu${}^{11}$,
T.~Anticic${}^{13}$,
N.~Antoniou${}^{18}$,
A.~G.~Asryan${}^{15}$,
B.~Baatar${}^{9}$,
A.~Blondel${}^{22}$,
J.~Blumer${}^{5}$,
L.~Boldizsar${}^{10}$
% ...

\vspace{1cm}

\noindent ${}^{ 1}$Cape Town University, Cape Town, South Africa \\
${}^{ 2}$ETH, Zurich, Switzerland \\
${}^{ 3}$Fachhochschule Frankfurt, Frankfurt, Germany \\
${}^{ 4}$Faculty of Physics, University of Sofia, Sofia, Bulgaria \\
${}^{ 5}$Forschungszentrum Karlsruhe, Karlsruhe, Germany \\
${}^{ 6}$Institute for Nuclear Research, Moscow, Russia \\
${}^{ 7}$Institute for Particle and Nuclear Studies, KEK, Tsukuba,  Japan \\
${}^{ 8}$Jagellionian University, Cracow, Poland  \\
${}^{ 9}$Joint Institute for Nuclear Research, Dubna, Russia \\
${}^{10}$KFKI Research Institute for Particle and Nuclear Physics, Budapest, Hungary \\
${}^{11}$LPNHE, University of Paris VI and VII, Paris, France \\
${}^{12}$Pusan National University, Pusan, Republic of Korea \\
${}^{13}$Rudjer Boskovic Institute, Zagreb, Croatia \\
${}^{14}$Soltan Institute for Nuclear Studies, Warsaw, Poland \\
${}^{15}$St. Petersburg State University, St. Petersburg, Russia \\
${}^{16}$State University of New York, Stony Brook, USA \\
${}^{17}$\'Swi{\,e}tokrzyska Academy, Kielce, Poland \\
${}^{18}$University of Athens, Athens, Greece \\
${}^{19}$University of Bergen, Bergen, Norway \\
${}^{20}$University of Bern, Bern, Switzerland \\
${}^{21}$University of Frankfurt, Frankfurt, Germany \\
${}^{22}$University of Geneva, Geneva, Switzerland \\
${}^{23}$University of Warsaw, Warsaw, Poland \\
${}^{24}$Warsaw University of Technology, Warsaw, Poland  \\

\newpage

\section{Introduction}
\label{intro}
The physics program of the NA61 (SHINE) (SHINE = SPS Heavy Ion and Neutrino Experiment) experiment at CERN SPS consists of three subjects (see references: \cite{intro1,intro2,intro3,run2007} for details).\\
In the first stage of data taking (2007-2009) measurements of hadron production in proton - nucleus interactions needed for neutrino (T2K) and cosmic-ray (Pierre Auger and KASCADE) experiments is performed. \\
In the second stage (2009-2010) hadron production in proton - proton and proton - nucleus interactions as a reference data for a better understanding of nucleus - nucleus reactions will be studied. \\
In the third stage (2009-2013) energy dependence of hadron production properties will be measured in nucleus - nucleus collisions as well as proton - proton, proton - lead interactions. The aim is to identify the properties of the onset of deconfinement and find evidence for the critical point of strongly interacting matter.\\
The experiment was approved at CERN in June 2007. The first pilot run was performed during October 2007.
The aims of this run were \cite{intro2}:
\begin{itemize}
\item to set up and test the NA61 (SHINE) apparatus,
\item to take pilot physics data on the interactions of \mbox{30 GeV} protons on two carbon targets with different geometry.
\end{itemize}
This article reports on the data taken in 2007 October run for the T2K neutrino oscillation experiment.
\section{The T2K experiment}
\label{sec:1}
The T2K experiment will study neutrino oscillations using off-axis neutrino beam from the \mbox{J-PARC} accelerator and the Super-Kamiokande detector \cite{aref,arefa}. The first phase of the T2K experiment (2009-2014) is aimed at:
\begin{itemize}
\item an order of magnitude better determination of the atmospheric parameters \mbox{$\Delta \sin^{2}2\theta_{23}=0.01$} and  \mbox{$\Delta m_{23}^{2}=10^{-4} eV^{2}$} by measuring the  \mbox{$\nu_{\mu} \to \nu_{x}$}  disappearance,
\item search for the $\nu_{\mu} \to \nu_{e}$ oscillations with the sensitivity to $sin^{2}2\theta_{13}$ down to 0.008 (90\%CL).
\end{itemize}

\begin{figure}
\begin{center}
\resizebox{0.80\textwidth}{!}{%
\ifpdf
\includegraphics{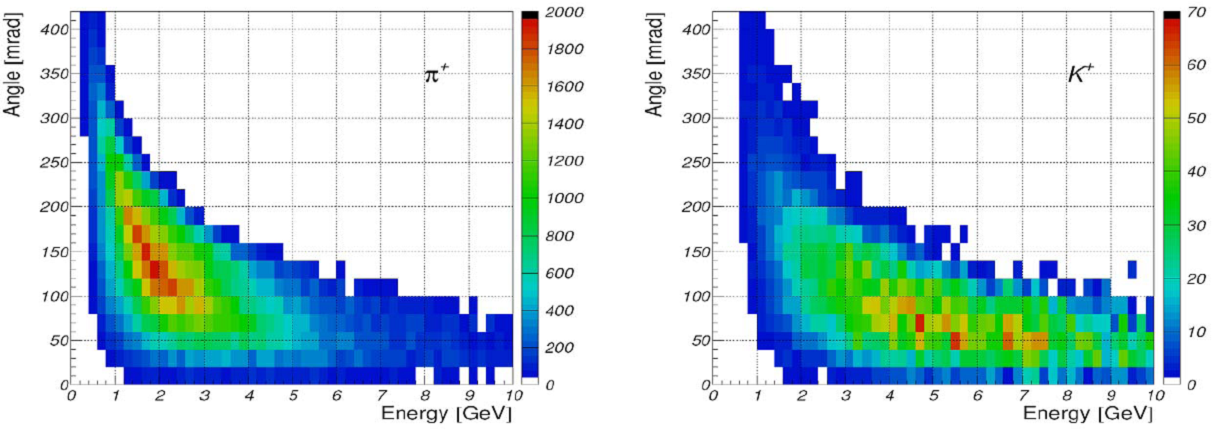}
\fi
}
\end{center}
\caption{Simulation of J-PARC neutrino beam. The plots show meson production angles versus energies for pions and kaons whose daughter neutrinos pass through the Super-Kamiokande detector (see \cite{ken}).}
\label{pion}       
\end{figure}

The beam neutrinos come from decays of pions and kaons produced in the interactions of the \mbox{30 GeV} protons on a carbon target (see Fig.~\ref{pion}). The neutrino interactions will be measured in the near detector (ND280) at a distance of \mbox{280 m} from the target and the Super-Kamiokande (SK) detector located at a distance of \mbox{295 km} from the neutrino source.
Both detectors are situated along the line \mbox{2.5 degrees} of the beam axis.\\

\begin{figure}[!h]
\begin{center}
\resizebox{0.80\textwidth}{!}{%
\ifpdf
\includegraphics{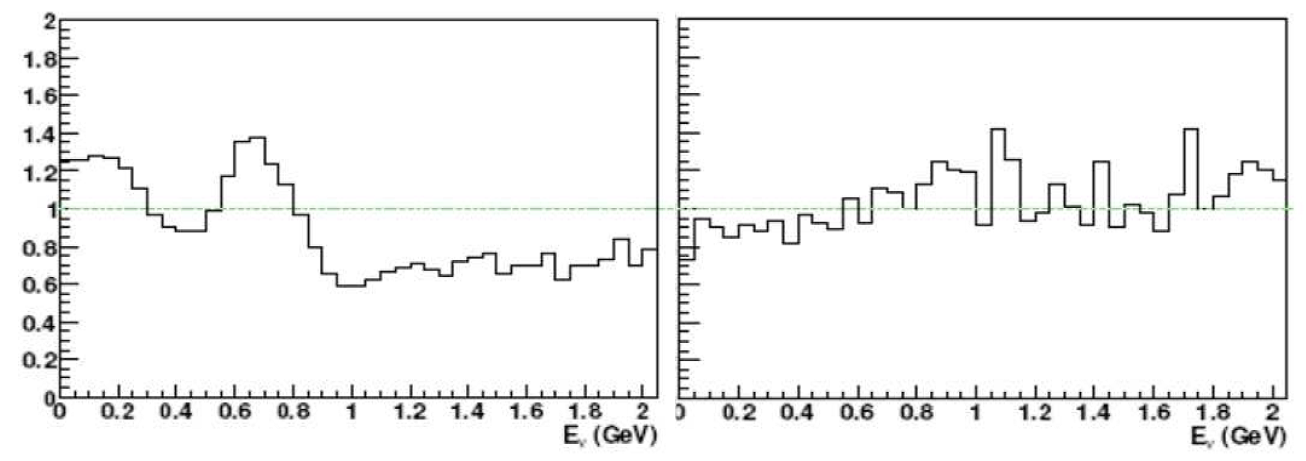}

\fi
}
\end{center}
\caption{Ratio of neutrino fluxes $R_{\nu_{\mu},\nu_{e}}$ for  $\nu_{\mu}$ and $\nu_{e}$, respectively. Clear energy dependence can be seen. Results are obtained from the T2K neutrino beam simulation.}
\label{r}       
\end{figure}

Neutrino oscillations will be probed by comparing observations at Super-Kamiokande with predictions with and without oscillations. The expected neutrino fluxes at SK, $\Phi^{SK}_{\nu_{e}}$ and $\Phi^{SK}_{\nu_{\mu}}$, will be obtained  from measured $\Phi^{ND}_{\nu_{e}}$ and $\Phi^{ND}_{\nu_{\mu}}$ at the near detector multiplied by the so-called far-to-near ratios, denoted $R_{\nu_{\mu}}$ and $R_{\nu_{e}}$ respectively:
\begin{equation}
\Phi^{SK}_{\nu_{\mu}, \nu_{e}} (E_{\nu})= R_{\nu_{\mu},\nu_{e}}(E_{\nu}) \cdot \Phi^{ND}_{\nu_{\mu}, \nu_{e}}(E_{\nu})
\end{equation}
 If the neutrino source was point-like and isotropic, the $R_{\nu_{\mu},\nu_{e}}$ ratios would be given by the ratio of the distances from the neutrino source squared and energy independent. \\However the neutrinos are born along the 94 m decay pipe which means that they constitute a point-like source only for the far detector. In addition near and far detectors see quite different solid angles. These effects lead to different energy spectra at the near and far detectors.\\
Knowledge of $R_{\nu_{\mu},\nu_{e}}$ is based only on Monte Carlo predictions (see Fig.~\ref{r}) in which many hadron production models are used. Studies show that these models may result in up to 20\% differences on the calculation of $R_{\nu_{\mu}}$ \mbox{\cite{intro2,nicolas}}. In the same reports it has been proved that to achieve T2K physics goals $R_{\nu_{\mu},\nu_{e}}$ should be known on the level of
\begin{center}
$\delta(R_{\nu_{\mu},\nu_{e}})\approx 2-3\%$ ,
\end{center}
which requires precise information on the pion and kaon production on the T2K target. For this purpose the  NA61 (SHINE) experiment has been proposed.

\section{The NA61 (SHINE) detector}
\begin{figure}
\begin{center}
\resizebox{0.90\textwidth}{!}{%
 \ifpdf
  \includegraphics{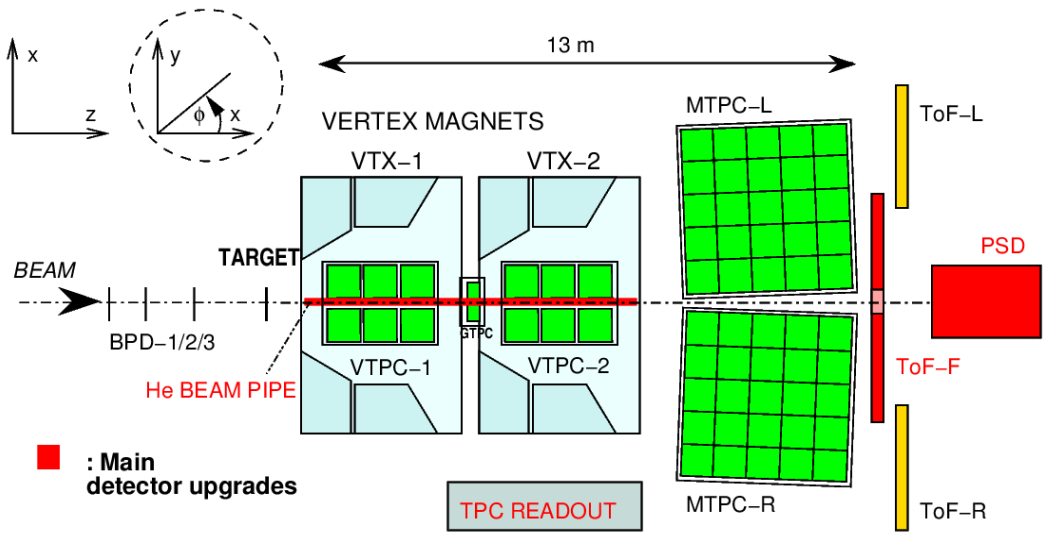}
 \fi
}
\end{center}
\caption{The layout of the NA61 (SHINE) set-up (top view, not in scale) with the basic upgrades indicated in red.}
\label{detektor}       
\end{figure}

The NA61 (SHINE) experiment is a large acceptance hadron spectrometer
at the CERN SPS. The layout of the NA61 (SHINE) set-up is shown in
Fig.~\ref{detektor}. The main components of the current detector
were constructed and used by the NA49 experiment \cite{detector}.
The main tracking devices are four large volume Time Projection Chambers (TPCs). Two of them, the vertex TPCs (\mbox{VTPC-1} and VTPC-2), are located in the magnetic field of two superconducting dipole magnets (1.5 and 1.1T, respectively).  Two others (MTPC-L and \mbox{MTPC-R}) are positioned downstream of the magnets symmetrically to the beam line. TPCs are filled with mixtures of $Ar+ CO_{2}$ (90:10) for VTPCs and (95:5) for MTPCs.
\\Two time-of-flight detectors (ToF-L/R) were inherited from NA49 and are able to provide a time measurement resolution of $\sigma \approx 100 ps$.
For the pilot 2007 run \cite{run2007} a new forward time-of-flight detector (ToF-F) was constructed in order to extend the acceptance of the NA61 (SHINE) set-up for pion and kaon identification as required for T2K measurements. The ToF-F wall is installed downstream of the MTPC-L and MTPC-R (red line on Fig.~\ref{detektor}), closing the gap between the ToF-L and ToF-R walls. The \mbox{ToF-F} time resolution is $\sigma \approx 120 ps$. The most downstream component of the NA61 (SHINE) apparatus is Projectile Spectator Detector (PSD) designed for heavy ion physics. It will be used to measure energy of the beam particles which do not interact with nucleons in the target.

\section {Beam counters and Beam Position Detectors}

NA61 is located in the North experimental area on the H2 beam line. The primary proton beam is extracted from the SPS towards the North area. Proton beam together with the production target creates secondary hadron beam which is coming to the H2 beam line.
The secondary beam contains a mixture of particles types: 87.3\% pions, 14.1\% protons and 1.6\% kaons.

\begin{figure}[!hb]
\begin{center}
\resizebox{0.80\textwidth}{!}{
 \ifpdf
  \includegraphics{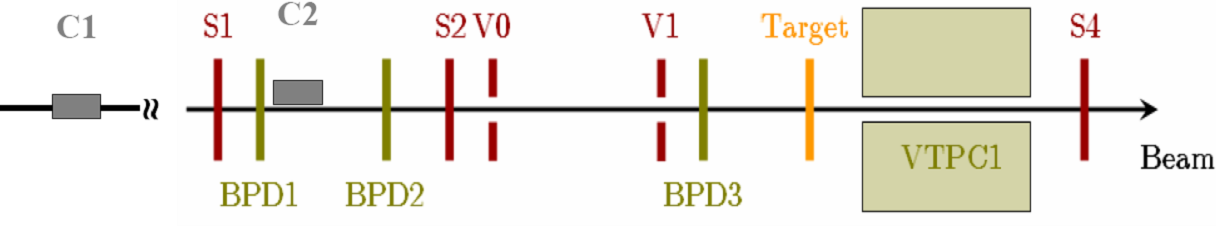}
 \fi
}
\end{center}
\caption{Schematic layout of the beam counters together with the three Beam Position Detectors (BPD).}
\label{beam}     
\end{figure}

The purpose of the beam instrumentation (see Fig.~\ref{beam}) is to tag the presence of the valid beam particle and to provide a precise timing information. The S1 counter is placed \mbox{32 m} upstream from the target. This counter defines the timing of the experiment. For proton beam a \mbox{5 mm} thick scintillator equipped with four photomultipliers was used. The S2 is additional scintillator.
A differential Cherenkov counter (C1) and threshold Cherenkov counter (C2) are applied to select protons from the secondary hadron beam. The beam definition is completed by V0 and V1, two scintillators with a hole, required to be in anticoincidence with S1, S2 and C1 and C2. These counters allow to select beam particles hitting the target.\\
The transverse positions of the incoming beam particles are measured by three Beam Position Detectors (BPDs). These counters are proportional chambers (48x48 $mm^{2}$) with cathode strip readout. The chambers are operated with an $Ar+ CO_{2}$ mixture (90:10).\\
The interactions in the target are selected by anticoincidence of the incoming beam particle with small($\phi=2cm$) scintillation counter (S4) placed on the beam line between two vertex magnets.

\section{ Carbon targets used in 2007 run}

Two carbon, isotropic graphite targets were used during 2007 run:
\begin{itemize}
\item a 2cm long target (about 4\% of nuclear interaction length, $\lambda_{I}$) with density $\rho = 1.84 \frac{g}{cm^{3}}$, so called thin target,
\item a 90 cm long cylinder (about 1.9 $\lambda_{I}$) of \mbox{2.6 cm} diameter, so called T2K replica target, with density \\
\mbox{$\rho=1.83 \frac{g}{cm^{3}}$}.
\end{itemize}

\vspace{2mm}

In 2007 run about \mbox{670 k} events with the thin target, \mbox{230 k} events with the T2K replica target and \mbox{80 k} events without target (empty target events) were reconstructed.\\
The data from the T2K replica target will allow to predict the $\nu$ flux of the T2K experiment. They will be used also for studies of secondary interactions in the target.\\
The thin target will be used for the determination of inclusive cross sections for the reactions:\\
\begin{center}
$p+C \to \pi^{+}(\pi^{-}) + X $,  \\
$p+C \to K^{+}(K^{-}, K^{0}_{s}) + X$.
\end{center}

\section{Event selection}

\begin{table}
\begin{center}

\begin{tabular}{llll}
\hline\noalign{\smallskip}
Cut & & &Result  \\
\noalign{\smallskip}\hline\noalign{\smallskip}
BPD position &  & &78\% \\
\textit{vertex.iflag = 0} & & &60\% \\
z-vertex position &  &  &41\% \\
\noalign{\smallskip}\hline
\label{table}
\end{tabular}
\end{center}
\caption{Preliminary results of event cuts applied on all thin target data. The fraction of events passing the cuts is presented.}
\end{table}

A proper event selection is needed in order to reduce the background coming from non-target interactions.
A preliminary list of the event cuts for the thin target data together with their impact on the event sample is presented in the Table~\ref{table}. 
The first cut requires that the particle was measured by each of the BPDs. The aim of this cut is to select well defined beam tracks. The second variable checks the flag set in each event during reconstruction. If the fit to determine the z-position of the main vertex has converged the \textit{vertex.iflag} is set to 0. In case of any problems with the fit the flag is set to non-zero values. The third position in the Table~\ref{table} requires that the fitted z-vertex position is close to the nominal one.

\section{Methods of Particle Identification}
In NA61 (SHINE) experiment particle identification is possible using
ionization measurements in active volume of the TPCs and time-of-flight measurements in ToF counters.\\

The energy loss distribution is described by the Landau distribution. Therefore in order to avoid large fluctuations in energy deposits the truncated mean technique is used in quantifying $\frac{dE}{dx}$. In this method, the highest and lowest measurements are rejected for each track. The accepted measurements follow approximately Gaussian distribution.
For all data collected during run in 2007 the (0:50) truncation was applied, which means that only the 50\% smallest clusters are kept for the determination of the $\frac{dE}{dx}$.

The optimization of the parameters required for the determination of energy loss in the TPCs was performed using the method developed by NA49 and described in detail in \cite{pp,veres}.
Corrections for the following effects were applied during the
calibration of the NA61 data:
\begin{itemize}
\item signal loss due to threshold cuts; corrections were obtained using Monte Carlo calculations for gas mixture used in NA61,
\item time dependence of the TPC gas pressure,
\item residual time dependence of the measurements \\(day/night),
\item charge absorption during the drift,
\item differences in the TPC  sector gain factors,
\item differences in the amplification of the preamplifiers and edge effects at sectors boundaries.
\end{itemize}

\begin{figure}
\begin{center}
\resizebox{0.90\textwidth}{!}{
 \ifpdf
  \includegraphics{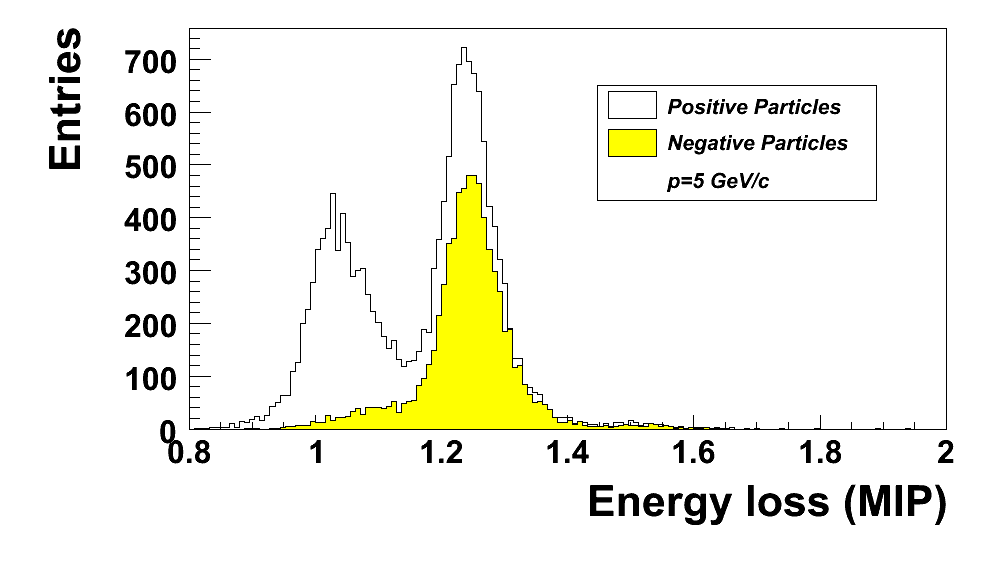}
  \includegraphics{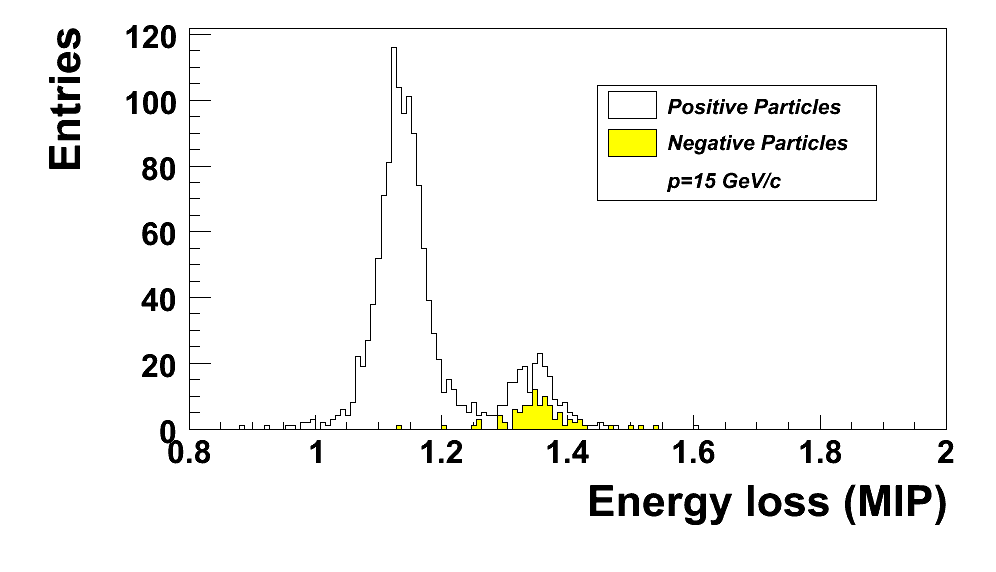}
 \fi
}
\end{center}
\caption{Energy loss for positively (white) and negatively (yellow) charged particles with momentum around 5 and \mbox{15 GeV/c}.}
\label{dedx}     
\end{figure}
\begin{figure}
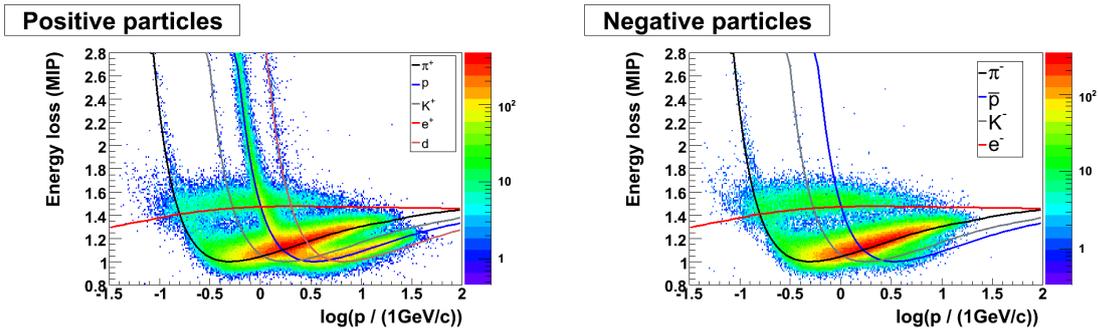

\begin{center}
\resizebox{0.96\textwidth}{!}{%
 \ifpdf
  \includegraphics{dedx_positive}
  \includegraphics{dedx_negative}
 \fi

}
\end{center}
\caption{Preliminary plots of the reconstructed dE/dx values versus momenta for positively  and negatively charged particles, respectively, together with the Bethe-Bloch curves for positrons (electrons), pions, kaons, (anti)protons and deuterons.}
\label{dedx1}       
\end{figure}

Preliminary distributions of corrected dE/dx values for particles  with momentum chosen around 5 and 15 GeV/c are presented in Fig.~\ref{dedx}. Estimated resolution of dE/dx measurements of 4-5\% can be achieved for particles passing through both the vertex and the main TPC chambers.\\
Scatter plots of the energy loss $\frac{dE}{dx}$ value of the track (in MIP units) versus the particle momentum in the laboratory frame is shown in Fig.~\ref{dedx1} for positively and negatively charged particles.\\
Here we present preliminary results obtained with the thin target.

\begin{figure}[!h]
\begin{center}
\resizebox{0.65\textwidth}{!}{
 \ifpdf
  \includegraphics{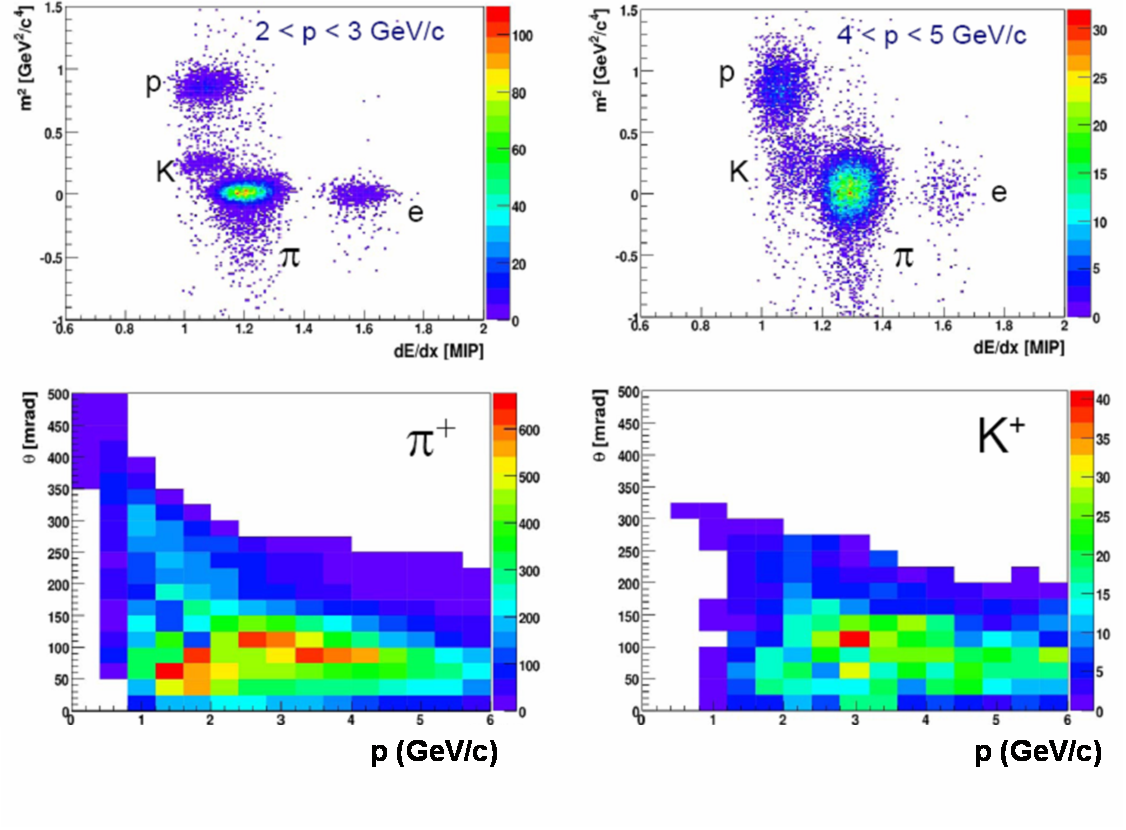}
 \fi
}
\end{center}
\caption{Preliminary results of  the combined dE/dx and ToF analysis in selected particle momentum bins.}
\label{dedx_tof}

\end{figure}

For particles reaching the ToF detectors dE/dx information
from the TPCs is available together with the time-of-flight.
Example of combined analysis is demonstrated on preliminary plots in Fig.~\ref{dedx_tof},  where particles separation in various momentum ranges is possible thanks to dE/dx signal
and the mass squared obtained from the forward ToF counters.\\
Comparison of   Fig.~\ref{pion} and Fig.~\ref{dedx_tof} (lower) proves that $\pi^{+}$ and $K^{+}$ detected in NA61 (SHINE) apparatus have momenta and production angles in the same range as needed for T2K neutrino oscillation experiment.\\
All presented plots show only raw results, without any corrections applied for acceptance, reconstruction efficiency or trigger normalization.

\section{Conclusions}
In the 2007 NA61 (SHINE) run the first physics data on interactions of \mbox{30 GeV} protons on the thin and T2K replica carbon target were registered. The NA61 (SHINE) apparatus together with new ToF-F counters were running successfully. All collected data were reconstructed and calibrated in order to obtain preliminary uncorrected particle spectra presented in this article.\\
In 2008 the NA61 (SHINE) Collaboration planned to increase the statistics of data needed for T2K neutrino oscillation experiment. The data taking was planned to last from September 4 to October 30. Unfortunately, it was stopped on October 6, because of the changes in accelerator schedule caused by the LHC incident \cite{memo}. During this shorter run installation and commissioning of the new TPC read-out electronics together with new DAQ (Data Acquisition) was successfully performed. This upgrade permits an increase of the data rate by a factor of 10 in comparison to the NA61 rate from 2007 year run. \\
Due to early run stop the physics data with the full detector upgrade were not registered. Thus, during 2009 run three weeks of data taking for T2K experiment are planned.

\section{Acknowledgment}
This work was supported by the Polish Ministry of Science and Higher Education (grant N N202 3956 33), the Hungarian Scientific Research Fund (OTKA 68506), the Virtual Institute VI-146 of Helmholtz Gemeinschaft, Germany, Korea Research Foundation (KRF-2008-313-C00200), the Federal Agency of Education of the Ministry of Education and Science of the Russian Federation (grant RNP 2.2.2.2.1547) and the Russian Foundation for Basic Research (grant 08-02-00018), the Ministry of Education, Culture, Sports, Science and Technology, Japan, \mbox{Grant-in-Aid} for Scientific Research (18071005, 19034011, 19740162), Swiss Nationalfonds Foundation 200020-117913/1 and ETH Research Grant TH-01 07-3.


\begin{thebibliography}{}
%
%
\bibitem{intro1}
N.~Antoniou {\em et al.} [NA61 Collaboration], CERN-SPSC-2006-034
\bibitem{intro2}
N.~Antoniou {\em et al.} [NA61 Collaboration], CERN-SPSC-2007-004
\bibitem{intro3}
N.~Abgrall {\em et al.} [NA61 Collaboration], CERN-SPSC-2007-019.
\bibitem{run2007}
N.~Abgrall {\em et al.} [NA61 Collaboration], CERN-SPSC-2008-018.
\bibitem{aref}
Y.~Itow {\em et al.}, LOI for JHF-nu experiment, \mbox{hep-ex/0106019}.
\bibitem{arefa}
Y. Yamada, Nucl. Phys. B \textbf{155}, (2006) 28.
\bibitem{ken}
K.~Sakashita, "Impact of the NA61 measurements on T2K results", [NA61 Collaboration Meeting],
http://indico.cern.ch/conferenceDisplay.py?confId=14045
\bibitem{nicolas}
N.~Abgrall, AIP Conf. Proc. \textbf{981} (2008) 157-159
\bibitem{detector}
S.~Afanasev {\em et al.} [NA49 Collaboration], Nucl. Instrum. Meth. \textbf{A 430}, (1999) 210 .
\bibitem{pp}
C.~Alt {\em et al.} [NA49 Collaboration], Eur. Phys. J. \textbf{C45}, (2006) 343.
\bibitem{veres}
G.~Veres, PhD thesis: \textit{"Baryon Momentum Transfer in Hadronic and Nuclear Collisions at the CERN NA49 Experiment"}, (2001)
\bibitem{memo}
M.~Gazdzicki, [NA61 Collaboration], CERN-SPSC-2008-026.
\end{thebibliography}
%

\end{document}